# Interferometric apodization by homothety – I. Optimization of the device parameters

J. Chafi,[1]★ Y. El Azhari,[1,2]★ O. Azagrouze,[1,2] A. Jabiri,[1] Z. Benkhaldoun ,[1,3] A. Habib[1,2] and Y. Errazzouki[1]

[1]*LPHEA, Faculté des Sciences Semlalia, Université Cadi Ayyad, Av. Prince My Abdellah, BP 2390, 40000 Marrakech, Morocco*
[2]*Centre Régional des Métiers de l'Education et de la Formation de Marrakech, 40000 Marrakech, Morocco*
[3]*Oukaimeden Observatory, Cadi Ayyad University, 40273 Marrakech, Morocco*

**ABSTRACT**
This study is focused on the very high dynamic imaging field, specifically the direct observation of exoplanetary systems. The coronagraph is an essential technique for suppressing the star's light, making it possible to detect an exoplanet with a very weak luminosity compared to its host star. Apodization improves the rejection of the coronagraph, thereby increasing its sensitivity. This work presents the apodization method by interferometry using homothety, with either a rectangular or circular aperture. We discuss the principle method, the proposed experimental set-up, and present the obtained results by optimizing the free parameters of the system while concentrating the maximum of the light energy in the central diffraction lobe, with a concentration rate of 93.6 per cent for the circular aperture and 91.5 per cent for the rectangular geometry. The obtained results enabled scaling the various elements of the experiment in accordance with practical constraints. Simulation results are presented for both circular and rectangular apertures. We performed simulations on a hexagonal aperture, both with and without a central obstruction, as well as a segmented aperture similar to the one used in the Thirty Meter Telescope (TMT). This approach enables the attainment of a contrast of approximately $10^{-4}$ at small angular separations, specifically around $1.8\lambda/D$. When integrated with a coronagraph, this technique exhibits great promise. These findings confirm that our proposed technique can effectively enhance the performance of a coronagraph.

## 1 INTRODUCTION

The direct observation of exoplanets faces a significant challenge due to the considerable difference in brightness between the planet and its host star. The brightness ratio is approximately $10^6$ in the infrared and can reach $10^{10}$ in the visible spectrum (Ridgway 2004). This renders direct observation of such objects virtually impossible. Nevertheless, the scientific community's efforts in the field of high dynamics have provided several approaches to overcome this difficulty.

The apodization method overcomes these challenges by diminishing the wings of the diffracted structure and concentrating the stellar light in the main diffraction lobe. The opportunities presented by apodization have prompted numerous scientific teams worldwide to propose various techniques for its implementation. Consequently, it can improve the efficiency of coronagraphs (Aime, Soummer & Ferrari 2001) by reducing noise caused by the secondary lobes of instrumental diffraction.

The apodized pupil Lyot coronagraph is one of the earliest techniques conceived in the early 2000s (Aime, Soummer & Ferrari 2002). It has been the subject of numerous papers (Soummer, Aime & Falloon 2003; Soummer 2005; Soummer et al. 2009, 2011). Additional techniques have been developed to broaden the array of solutions (Por et al. 2020; Will & Fienup 2020). At present, the vector-apodizing phase plate (vAPP) is a promising technique that could be incorporated into the current generation of coronagraphs (Snik et al. 2012; Bos et al. 2021).

In contrast to the approach using an amplitude transmission mask, interferometric apodization significantly reduces diffraction effects (Aime et al. 2001). In the case of a rectangular aperture, the process occurs in two stages. Each stage can perform apodization along one of the two symmetry axes. For each axis, the process involves dividing the amplitude of the focal image and then coherently recombining the two images after suitable shifting (Aime et al. 2002; Aime 2005). El Azhari et al. (2005) and Azagrouze, El Azhari & Habib (2008) have carried out experimental apodization using both a Michelson interferometer (MI) and rectangular aperture, while Carlotti et al. (2008) has employed a Mach–Zehnder interferometer (MZI) to achieve one-dimensional apodization.

The scientific literature has extensively studied the use of coronagraphs for high-resolution imaging of planetary systems. Over the years, different families of coronagraphs have been developed to improve instrument performance. Among these coronagraphs, the apodized pupil Lyot coronagraph (APLC) has gained popularity due to its ease of implementation and high contrast capabilities. Several recent research works have focused on the optimization of the APLC to further improve its performance (N'Diaye, Pueyo & Soummer 2015; N'Diaye et al. 2016, 2018; St. Laurent et al. 2018, 2019), the use of hexagonal apertures (Nickson et al. 2022), and the use of phase instead of amplitude, PAPLC, (Por 2020; Por et al. 2023).

★ E-mail: chafi.jamal2@gmail.com (JC); youssef.elazhari@men.gov.ma (YEA)

Other types of coronagraphs, such as phase-induced amplitude apodization complex mask coronagraphs (Guyon et al. 2010), are designed to be used with large segmented apertures placed on-axis, such as the one proposed for the large ultraviolet optical infrared surveyor (LUVOIR-A; Sirbu et al. 2021). A laboratory demonstration was conducted with a segmented aperture by Marx et al. (2021), as well as with a high-contrast obstructed segmented aperture by Belikov et al. (2022). Vortex coronagraphs (Foo, Palacios & Swartzlander 2005; Mawet et al. 2005) have been extensively researched in the scientific literature, and have shown promising results for high-contrast imaging of exoplanets. In particular, there has been recent research on combining vortex coronagraphs with new types of apodization for use in large space missions such as LUVOIR (Fogarty et al. 2017; Ruane et al. 2018).

It is important to note that another type of apodization is utilized in conjunction with Lyot-type coronagraphs for large space missions, with the goal of enhancing the performance of both APLC and vortex coronagraphs and achieving an average contrast of $10^{-10}$ in the upcoming LUVOIR telescope (Stahl, Shaklan & Stahl 2015; Stahl 2017; Leboulleux et al. 2017, 2018a, b; Laginja et al. 2019, 2020, 2021; The LUVOIR Team 2019; Stahl, Nemati & Stahl 2020). Modelling and performance analysis studies have also been conducted in this regard (Juanola-Parramon et al. 2022).

None the less, this technique does not work with circular apertures similar to those of telescopes. To address this challenge, Carlotti et al. (2008) carefully positioned phase plates to achieve apodization with a circular pupil using a Mach–Zehnder interferometer set-up. Another solution proposed by Azagrouze et al. (2010) and Habib et al. (2010), called interferometric apodization by homothety (IAH), involves splitting a parallel beam of light from the telescope's focal plane and controlling the width and intensity of each of the two beams before interfering them.

In this study, we focus on interferometric apodization by homothety, which enhances the coronagraph's ability to reject stellar light. The IAH apodization technique is an effective method for achieving very high dynamic range (HDR) imaging by reducing diffraction noise in final images. This attenuation of diffracted stellar brightness allows for an increase in the resolution of the final image, making it easier to discern fine details in the surrounding structures. The IAH technique is intended to be used in combination with a coronagraph to enhance its performance. One of the many advantages of this technique is its simplicity of implementation, as it does not require complex or expensive equipment and integrates easily with existing coronagraphs. The method also lends itself well to cascading multiple IAH stages: the input of stage $n$ is placed at the output of stage $n-1$, allowing for further improvement of the point spread function (PSF) sidelobe rejection rate. Additionally, it uses simple achromatic optical elements such as apertures and neutral density filters, facilitating the assembly of the apodizer and reducing chromatic aberrations. The IAH technique offers a relatively wide range of values for the optimization parameters ($\gamma$ and $\eta$), allowing for modulation of the attenuation zone and potential detection of exoplanets based on their angular distance. Overall, the IAH technique is a powerful tool for achieving HDR imaging and improving the performance of existing coronagraphs.

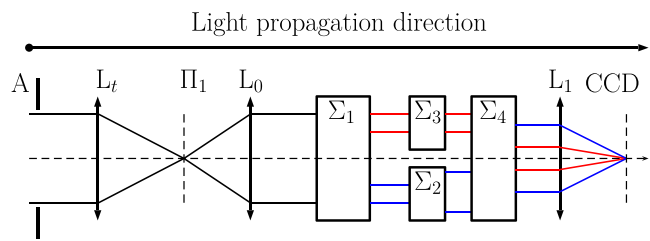

**Figure 1.** Principle of interferometric aperture apodization using homothety. $A$ and $L_t$ simulate the telescope. $\Pi_1$ is the telescope's focal plane. $\Sigma_1$ separates the beam into two coherent beams. $\Sigma_2$ stretches out one of the beams by applying a homothety with coefficient $\eta$. $\Sigma_3$ compensates for a possible phase shift introduced by $\Sigma_2$. $\Sigma_4$ connects the beams at the output from the system. The interference result is observed in the focal plane of $L_1$ where the *CCD* sensor is placed.

Ongoing simulation work in our laboratory involves the application of the IAH technique on phase mask coronagraphs (4QPM, 8OPM, Roddier & Roddier, and Vortex) and Lyot amplitude mask coronagraphs. This work is set to be reported in an article currently under preparation. The initial numerical simulations suggest remarkable contrasts at very small angular separations.

The innovation of this study lies in a systematic analysis aimed at obtaining optimal and more accurate values of the parameters ($\gamma$ and $\eta$) for IAH compared to those presented in Habib et al. (2010). Furthermore, this research includes a sizing study of the assembly to facilitate the experimental implementation of the IAH technique. It also explores the application of apodization on a rectangular symmetric aperture to evaluate the effectiveness of the IAH technique and carries out simulations on hexagonal apertures, This method has facilitated achieving a contrast on the order of $10^{-4}$ at minimal angular separations, particularly around $1.8\lambda/D$, making it a promising technique when combined with a coronagraph. In Section 2, we describe the principle of the IAH apodization technique and present a proposed experimental set-up based on the Mach–Zehnder interferometer. Section 3 is devoted to a detailed study of the circular aperture, including the determination of optimal values for the free parameters and simulation results using standard values of the proposed optical components. In Section 4, we demonstrate that the apodization technique can be applied to a rectangular aperture and present the optimal values for the free parameters as well as simulation results using the standard characteristics of the optical components in the configuration. In Section 5, we demonstrate the feasibility of using the IAH apodization technique on hexagonal apertures, including a simple hexagonal aperture and a hexagonal aperture with central obstruction, as well as a segmented aperture similar to that used in the TMT telescope.

## 2 PRINCIPLE OF THE TECHNIQUE AND PROPOSED SET-UP

### 2.1 Principle of the technique

We reviewed the configuration of interferometric apodization using homothety with rectangular or circular apertures. The concept involves replicating the initial PSF to obtain two similar PSFs using amplitude division (Habib et al. 2010). A coherent superposition is then performed to reduce diffraction wings. The preferred approach involves stretching one of the PSFs and controlling its amplitude to match the negative lobes of one of the two PSFs with the positive lobes of the other (see Fig. 1).

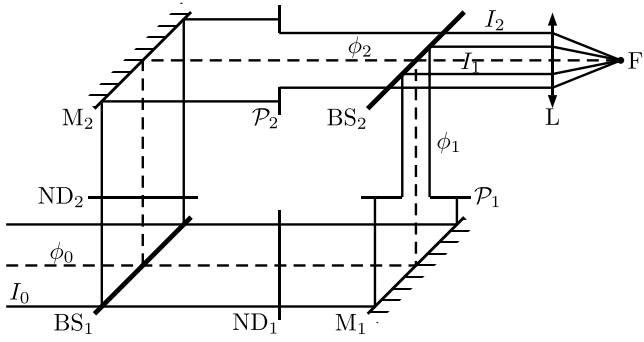

**Figure 2.** Schematic diagram of the set-up, was designed to work with both single diffraction holes or rectangular aperture, depending on the type of apodization desired, proposed for the realization of interferometric apodization by homothety using a Mach–Zehnder interferometer. $\mathcal{P}_1$ and $\mathcal{P}_2$ are two diffraction holes of diameters $\phi_1$ and $\phi_2$, respectively (or two rectangular aperture of sides $x_1$, $y_1$ and $x_2$, $y_2$, respectively). ND$_1$ and ND$_2$ denote two neutral density filters. $\phi_0$ is the diameter of the incident beam of intensity $I_0$.

## 2.2 Proposed experimental set-up

The experimental device that we propose (Fig. 2) was designed to work with both circular or rectangular pupils, depending on the type of apodization desired, and two neutral density filters judiciously placed in the two arms of an MZI. The interferometer ensures the separation–recombination of PSF, while the diffraction holes (or rectangular apertures) allow to control their diameter and the neutral filters their intensity. Fig. 2 shows the schematic representation of such a set-up. $\mathcal{P}_1$ and $\mathcal{P}_2$ are the two circular holes of respective radii $\phi_1$ and $\phi_2$ (or two rectangular apertures of side $x_1$, $y_1$ and $x_2$, $y_2$, respectively) while ND$_1$ and ND$_2$ are the two neutral density filters of transmission factors in intensity $T_i = 10^{-D_i}$ ($i = 1, 2$). The beam splitters lames correspond to a 50 per cent transmission–reflection. The intensities of the two beams which are combined at the output of the interferometer are

$$I_1 = \frac{1}{4}\left(\frac{\phi_1}{\phi_0}\right)^2 T_1 I_0 \quad \text{and} \quad I_2 = \frac{1}{4}\left(\frac{\phi_2}{\phi_0}\right)^2 T_2 I_0. \quad (1)$$

The diameters of the diffraction holes and the densities of the neutral filters are linked by the apodization parameters $\gamma$ and $\eta$ such as

$$\begin{cases} \phi_2 = \eta\,\phi_1 \\ \Delta D = D_2 - D_1 = 2\log_{10}\left(\frac{\eta}{\gamma}\right) \end{cases}. \quad (2)$$

## 3 CIRCULAR APERTURE

In this section, we introduce the method of interferometric apodization for a circular aperture using homothety, and briefly review the general formalism and optimization results that guide the sizing of optical components. Additionally, we discuss the experimental configuration for implementation and present the outcomes of numerical simulations.

### 3.1 Formalism

In the case of a binary circular aperture with a perfectly planar incident wavefront, the PSF in the focal plane is given as follows (Born & Wolf 1980)

$$\Psi(u) = 2\frac{J_1(\pi u)}{\pi u}, \quad (3)$$

where $J_1$ is the first-order Bessel function. The variable $u$ is expressed as a function of the angular position $\theta$, diameter $D$, and wavelength $\lambda$, such that $u = D\,\theta/\lambda$, with $\theta = r/f$, where $r$ is the radial position in the focal plane and $f$ is the focal length. The intensity associated with the PSF in the focal plane is written as

$$I = I_0\left|2\frac{J_1(\pi u)}{\pi u}\right|^2, \quad (4)$$

where $I_0$ is the maximum intensity at the center of the Airy disc. The angular radius of the Airy disc is $1.22\,\lambda/D$. This central disc contains approximately 84 per cent of the total energy, with the remainder distributed across the secondary diffraction lobes.

The resulting amplitude of the beam recombination in the focal plane is expressed as the sum of two PSFs corresponding to aperture diameters $D$ and $\eta D$, respectively. It is written as

$$\Phi(u) = \Psi(u) + \gamma\,\Psi(\eta u) = 2\frac{J_1(\pi u)}{\pi u} + 2\gamma\frac{J_1(\pi u \eta)}{\pi u \eta}. \quad (5)$$

The real parameters $\gamma$ and $\eta$ in the last expression enable the optimization of the superposition between the two PSFs such that their sidelobes compensate for each other as much as possible, resulting in apodization.

### 3.2 Optimization

The more effective the apodization, the lower the light energy outside the central diffraction disc. This efficiency depends on the values of the parameters $\gamma$ and $\eta$, which can maximize the energy in the central disc. The optimization criterion used involves maximizing the percentage of energy in the central disc, which is equivalent to finding the maximum ratio

$$\epsilon(\gamma, \eta) = \frac{\int_0^{u_m(\gamma,\eta)} I(\gamma, \eta, u)\,2\pi\,r\,\mathrm{d}r}{\int_0^{\infty} I(\gamma, \eta, u)\,2\pi\,r\,\mathrm{d}r}, \quad (6)$$

where $u_m(\gamma, \eta)$ is the respective value of the first dark, with $I(\gamma, \eta, u_m) = 0$.

For a circular aperture, the intensity in the final focal plane is given by

$$I(\gamma, \eta, u) = I_0\left|2\frac{J_1(\pi u)}{\pi u} + 2\gamma\frac{J_1(\pi u \eta)}{\pi u \eta}\right|^2. \quad (7)$$

We have sought the values of $\gamma$ and $\eta$ that minimize the energy outside the central disc of radius $u_m(\gamma, \eta)$. In Fig. 3, we present the evolution of the ratio (equation 6) as a function of the parameters $\gamma$ and $\eta$. We observe a maximum of 93.6 per cent for the values $\gamma_{\max} = 0.525$ and $\eta_{\max} = 0.625$.

### 3.3 Sizing

In practice, the selection of components offered in optical equipment supplier catalogues is quite limited. As a result, it may not always be possible to find components (diffraction apertures and neutral density filters) that exactly match the optimal values determined in Section 3.2. It is then necessary to perform sizing work to get as close as possible to the optimal conditions while considering the available components offered. We began by choosing the value of $\phi_1 = 1\,000\,\mu$m, which is available in many optical component supplier catalogues. In this case, the optimal diameter value for the second aperture is approximately $625\,\mu$m. For neutral filter selection, the optimal value of the density difference $\Delta D$ is 0.15. The neutral filters that allowed us to approach this optimal value have the following densities: $D_1 = 0.20$ and $D_2 = 0.04$, resulting in a standard value of $\Delta D = 0.16$. Table 1 summarizes the values for standard components that we used in this work, and provides a comparison with the corresponding optimal values.

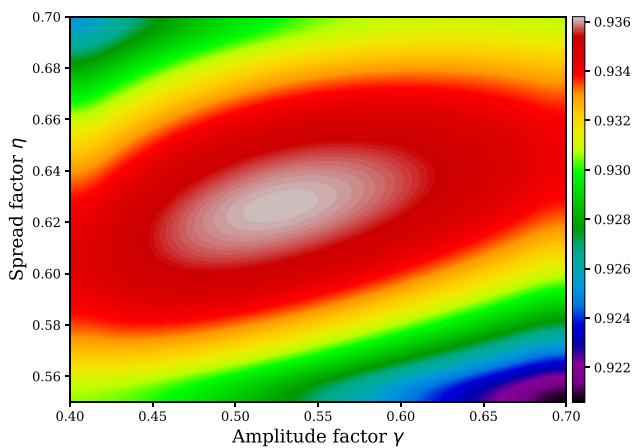

**Figure 3.** Evolution of the ratio $\epsilon$ as a function of the parameters $\gamma$ and $\eta$. The maximum of $\epsilon$ corresponds to the following values: $\gamma_{max} = 0.525$ and $\eta_{max} = 0.625$.

**Table 1.** Optimal and standard values of the diameters $\phi_1$ and $\phi_2$ of the diffraction apertures, the densities $D_1$ and $D_2$, as well as the interferometric apodization parameters $\gamma$ and $\eta$ are presented. 'Standard' specifies the values of components available in the catalogues of optical equipment suppliers.

| Configuration | $\phi_1(\mu m)$ | $\phi_2(\mu m)$ | $\eta$ | $\gamma$ | $\Delta D$ | $D_1$ | $D_2$ |
|---|---|---|---|---|---|---|---|
| Optimal | 1000 | 625 | 0.625 | 0.525 | 0.15 | – | – |
| Standard | 1000 | 600 | 0.600 | 0.498 | 0.16 | 0.20 | 0.04 |

### 3.4 Simulation results

In this section, we present the PSF simulations using both optimal and standard values. We will also use (equation 7) to plot the output intensity of the Mach–Zehnder interferometer. In Fig. 4, we plot the radial cuts of PSF intensities normalized on both linear and logarithmic scales (Figs 4a and b). For comparison, the Airy disc is shown in dotted lines. The red curve corresponds to apodization using the optimal values ($\gamma = 0.525$ ; $\eta = 0.625$), and the blue curve corresponds to the standard values ($\gamma = 0.498$ ; $\eta = 0.6$).

In the optimal case, Fig. 4 shows a significant reduction in diffraction side lobes compared to the Airy's diffraction pattern. We can see that 93.6 per cent of the total light energy is concentrated in the central PSF's disc, while in the Airy's disc case, this percentage is only 84 per cent. The energy in the secondary lobes is thus reduced. We also noticed that for $1.8\,\lambda/D < \theta < 3\,\lambda/D$ and from $4.5\,\lambda/D$, the attenuation of the secondary lobes is more significant. We have developed an interferometric apodization technique based on homothety (IAH), which allowed us to achieve a contrast of $10^{-4}$ at a distance very close to $2.5\,\lambda/D$, and a contrast of the order of $10^{-7}$ at a distance of $7.8\,\lambda/D$. The performance achieved with our IAH-based approach is comparable to that of other apodization techniques aiming to mitigate the effects of starlight diffraction. For instance, Otten et al. (2014) achieved a contrast on the order of $10^{-3.8}$ between 2 and $7\lambda/D$ while taking certain optical aberrations into account, while Por (2017) predicted a contrast of $10^{-5}$ at distances from 1.8 to $10\lambda/D$ using an apodizing phase plate. Similarly, Zhang et al.

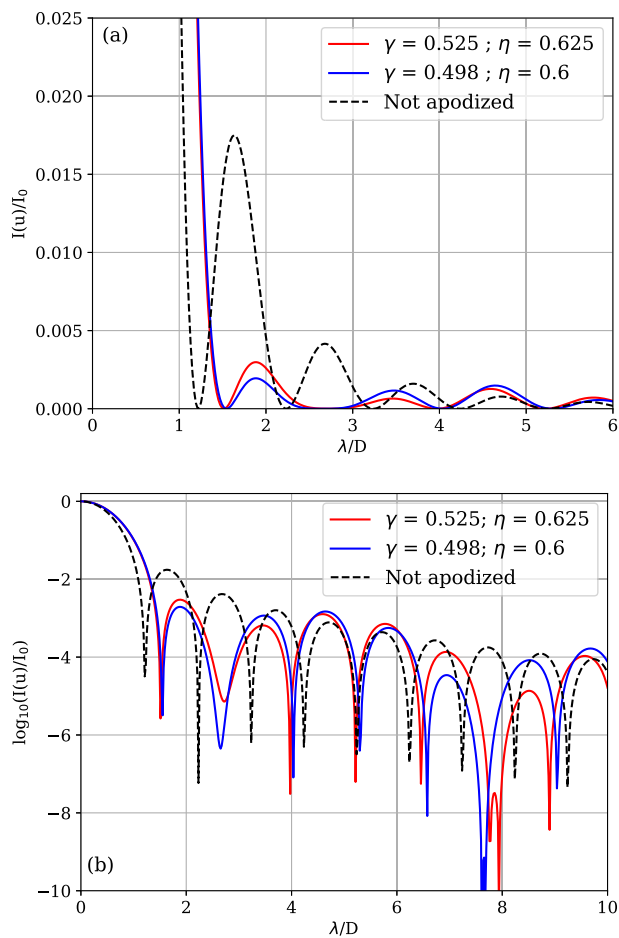

**Figure 4.** (a) Radial cross-sections of the PSF intensities normalized using linear scales and (b) normalized using logarithmic scales. The Airy's classical PSF is represented by the dotted line, while the red and blue lines represent the PSF apodized for $\gamma = 0.489$; $\eta = 0.6$ and $\gamma = 0.525$; $\eta = 0.625$, respectively.

(2018) utilized 'microdot apodizers' to suppress diffracted starlight, and Asmolova et al. (2018) explored 'apodized photon sieves' (APS) to achieve high contrast levels in all directions. These results allow us to anticipate promising performance for coronagraphs equipped with an IAH apodizer, this has motivated us to initiate a process of practical validation of these simulations. The outcomes of this experimental work will be published in a forthcoming article.

The results of simulations using the standard values for $\gamma$ and $\eta$ parameters (instead of optimal values) also show a significant reduction in diffraction lobes (see Fig. 4). This can be explained by the fact that standard values were chosen to be as close as possible to optimal values, and that the $\epsilon(\gamma, \eta)$ curve has a relatively flat section around its maximum (see Fig. 3).

Fig. 5 illustrates the impact of IAH in 2D for the optimal scenario ($\gamma = 0.525$; $\eta = 0.625$). The intensity of the star was simulated with and without apodization, and a planet with an angular distance of $1.8\,\lambda/D$ and relative intensity of $10^{-4}$ was added (shown in Figs 5c and d). The effect of apodization is evident, as it reduces the annuli and enhances the planet's visibility. Fig. 6 demonstrates the same phenomenon using standard values ($\gamma = 0.498$; $\eta = 0.6$). Once again, apodization enables the exoplanet with a relative intensity of $10^{-4}$ at an angular distance of $1.8\,\lambda/D$ to be distinguished from the star.

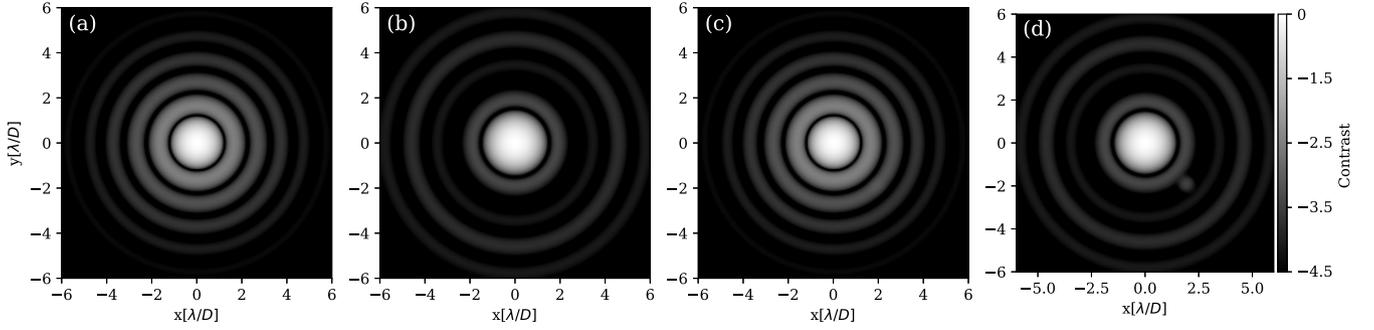

**Figure 5.** Normalized 2D intensity. (a) Star intensity without apodization, (b) star intensity with apodization, (c) star and planet intensity without apodization, and (d) star and planet intensity with apodization, with an intensity ratio between the star and the planet of $10^{-4}$ at an angular separation of $1.8\,\lambda/D$. Apodization with optimal values ($\gamma = 0.525$; $\eta = 0.625$). The colour bar indicates the level of contrast for all images a, b, c, and d.

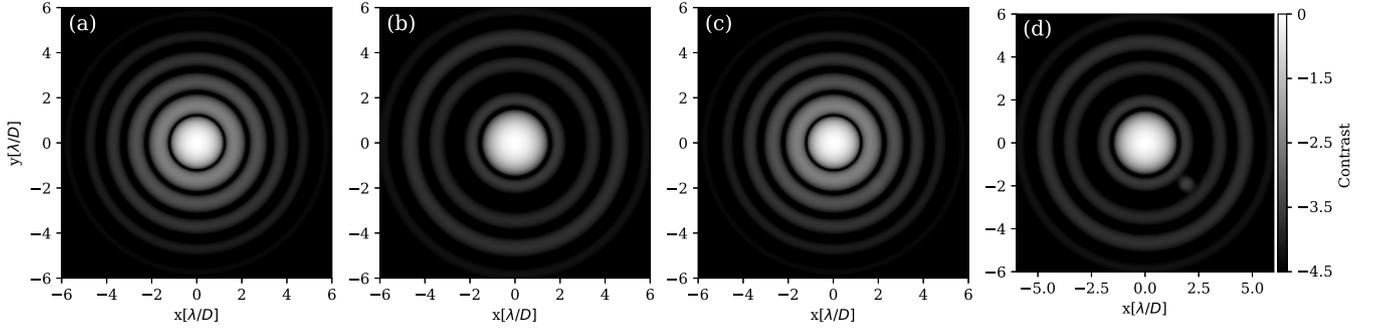

**Figure 6.** Normalized 2D intensity. (a) Intensity of the star without apodization, (b) intensity of the star with apodization, (c) intensity of the star and planet without apodization, and (d) intensity of the star and planet with apodization, with an intensity ratio between the star and the planet of $10^{-4}$ at an angular separation of $1.8\,\lambda/D$. Apodization with standard values ($\gamma = 0.498$; $\eta = 0.6$). The colour bar indicates the level of contrast for all images a, b, c, and d.

## 4 RECTANGULAR APERTURE

Interferometric apodization of a rectangular aperture is a method of apodization that employs an interferential process to achieve cosine transmittances. This approach was first proposed by Aime et al. (2001). We have selected rectangular apertures to study interferometric apodization for several reasons. First, rectangular geometry has been shown to be interesting by various publications such as Zanoni & Hill (1965) and Nisenson & Papaliolios (2001), as it presents a double apodization in the diagonal direction of the aperture. Furthermore, a recent study conducted by Itoh & Matsuo (2022) specifically focuses on the use of rectangular apertures to achieve deep nulls. Secondly, for the experimental validation of IAH results, we chose to begin with the study of rectangular apertures before moving on to circular apertures. This decision was made due to the ease of implementing rectangular apertures with high-quality components and the ability to vary the $\gamma$ parameter (adjustable aperture) easily. Furthermore, the study of rectangular apertures allows for a general and precise understanding of the performance of the IAH technique and its effectiveness before applying it to more complex telescope pupils, such as hexagonal pupils. We extend the approach of Habib et al. (2010) on homothetic apodization using a circular aperture to include the application of this technique to a rectangular aperture.

### 4.1 Basic formulas

For one-dimensional telescope of width $(a, b)$, the pupil $P(x, y)$ may be simply written as the window function $\Pi(\frac{x}{a}, \frac{y}{b})$. The pupil $P(x, y)$ is written as

$$P(x, y) = \begin{cases} \Pi\left(\frac{x}{a}, \frac{y}{b}\right) = 1 & \text{if } \frac{-a}{2} < x < \frac{a}{2} \text{ and } \frac{-b}{2} < y < \frac{b}{2} \\ 0 & \text{other} \end{cases} \quad (8)$$

In the focal plane, the amplitude of the monochromatic PSF is proportional to its Fourier transform, given by

$$\psi(u, v) = \frac{\sin(\pi u)}{\pi u} \frac{\sin(\pi v)}{\pi v}. \quad (9)$$

Here, $u = \frac{a\alpha}{\lambda}$ and $v = \frac{b\beta}{\lambda}$, where $\alpha$ and $\beta$ represent the observation direction, and $(a, b)$ denote the widths in the $(x, y)$ directions.

The intensity of the PSF in the focal plane can be expressed as

$$I(u, v) = I_0 \left| \frac{\sin(\pi u)}{\pi u} \frac{\sin(\pi v)}{\pi v} \right|^2. \quad (10)$$

Here, $I_0$ represents the maximum intensity at the central fringe.

The IAH technique, as shown in Fig. 1, is applicable to rectangular geometries. The coherent superimposition of two PSFs from a rectangular aperture, with one undergoing broadening and amplitude modulation by the ratio $\gamma$, leads to the ensuing expression for the ultimate image in the focal plane

$$\psi(\gamma, \eta, u, v) = \frac{\sin(\pi u)}{\pi u} \frac{\sin(\pi v)}{\pi v} + \gamma \frac{\sin(\pi u \eta)}{\pi u \eta} \frac{\sin(\pi v \eta)}{\pi v \eta}. \quad (11)$$

The real parameters $\gamma$ and $\eta$ in the amplitude expression (equation 11) can be adjusted to optimize the superposition between the two PSFs, leading to efficient apodization.

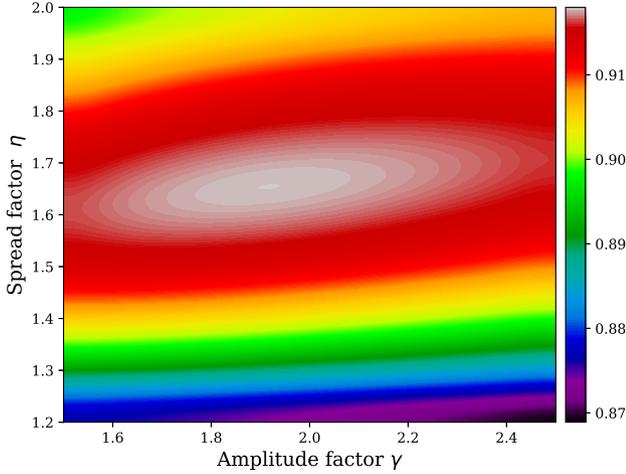

**Figure 7.** Evolution of the ratio $\epsilon$ as a function of the parameters $\gamma$ and $\eta$. The maximum of $\epsilon$ corresponds to the following values: $\gamma_{max} = 1.928$; $\eta_{max} = 1.657$.

### 4.2 Optimization

We have previously discussed the optimization process for achieving apodization efficiency in the case of a circular aperture in Section 3.2. In the case of rectangular apertures, we follow the same approach to attain optimal efficiency by adjusting the values of $\eta$ and $\gamma$ to achieve maximum concentration of energy in the central lobe, maximizing the ratio

$$\epsilon(\gamma, \eta) = \frac{\int_{-x_m}^{x_m} \int_{-y_m}^{y_m} I(\gamma, \eta, u, v)\, dx\, dy}{\iint_0^{+\infty} I(\gamma, \eta, u, v)\, dx\, dy}, \quad (12)$$

where $x_m$ and $y_m$ are the values of $x$ and $y$, at which the first dark $I(\gamma, \eta, x, y) = \psi^2(\gamma, \eta, u, v) = 0$. The intensity, for a rectangular aperture in the focal plane is given by

$$I(\gamma, \eta, u, v) = \left| \frac{\sin(\pi u)}{\pi u} \frac{\sin(\pi v)}{\pi v} + \gamma \frac{\sin(\pi u \eta)}{\pi u \eta} \frac{\sin(\pi v \eta)}{\pi v \eta} \right|^2. \quad (13)$$

Fig. 7 provides a graphical representation of the evolution of the ratio (equation 12) with respect to the parameters $\gamma$ and $\eta$. In this figure, $\epsilon$ attains a maximum value of $\epsilon_{max} = 91.8$ per cent for $\gamma_{max} = 1.928$ and $\eta_{max} = 1.657$. In contrast, in the unapodized case ($\gamma = 0$ and $\eta = 1$), the percentage of light intensity in the main peak is only 81.5 per cent. This indicates a gain of at least 10.3 per cent compared to the unapodized case.

### 4.3 Simulation results

This section is dedicated to simulating the case of a rectangular aperture. To achieve this, we utilize equation (13) to plot the intensity obtained at the output of the MZI. The radial cuts of the PSF in the final focal plane are presented in Fig. 8, in logarithmic scale, for both the unapodized case ($\gamma = 0$; $\eta = 1$) and the apodized case with optimal values $\gamma_{max} = 1.928$ and $\eta_{max} = 1.657$ (red dotted line).

It is evident that the diffraction side lobes are significantly reduced in the apodized case, and the diagonal intensity ($x = y$) on the right-hand side is enhanced. Thus, in this case, the effect of apodization is more pronounced. Furthermore, at a distance of 2, $5\lambda/D$, we can observe a contrast of the order of $10^{-4.5}$. If we move away from the optical axis at long distances ($\sim 7.5\lambda/D - 10\lambda/D$), the contrast will be even higher, on the order of $10^{-6.5}$.

Fig. 9 demonstrates the effect of interferometric apodization by homothety in 2D for the optimal case ($\gamma_{max} = 1.928$; $\eta_{max} = 1.657$). We simulated the intensity of the star in both cases, with and without apodization. Additionally, we included the planet intensity (Figs 9c and d) located at an angular distance of $2.5\lambda/D$ and with a relative intensity of $10^{-4.5}$ compared to the host star. It is evident that apodization reduces the annuli and enhances the planet's visibility.

## 5 APPLICATION ON OTHER PUPILS

Current and future telescopes are increasingly using hexagonal apertures due to their numerous advantages over circular apertures (Sabatke, Burge & Sabatke 2005; Soummer et al. 2009). Hexagonal segments offer more efficient light collection and a more compact design, which significantly reduces weight and manufacturing costs. Hexagonal apertures also have the advantage of reducing the loss of light due to diffraction, resulting in sharper images. Manufacturing and aligning hexagonal segments are also simpler than for circular segments. These advantages make hexagonal apertures the preferred choice for most current and future segmented mirror telescopes. Moreover, Sabatke et al. (2005) emphasized that apodizing a hexagonal pupil could offer the possibility of obtaining high-contrast images during space observation. For these reasons, we conducted simulations of the IAH technique on hexagonal pupils.

In this section, the results of the interferometric apodization simulation are presented. The simulation employs the homothety technique on two types of apertures: a simple and obstructed hexagonal aperture, as well as an aperture similar to that of the TMT telescope. The simulation was performed using optimization values for rectangular geometry, corresponding to the parameters $\gamma = 1.928$ and $\eta = 1.657$. However, an optimization study of $\gamma$ and $\eta$ values is necessary and will take several weeks, with the results being presented in a forthcoming article.

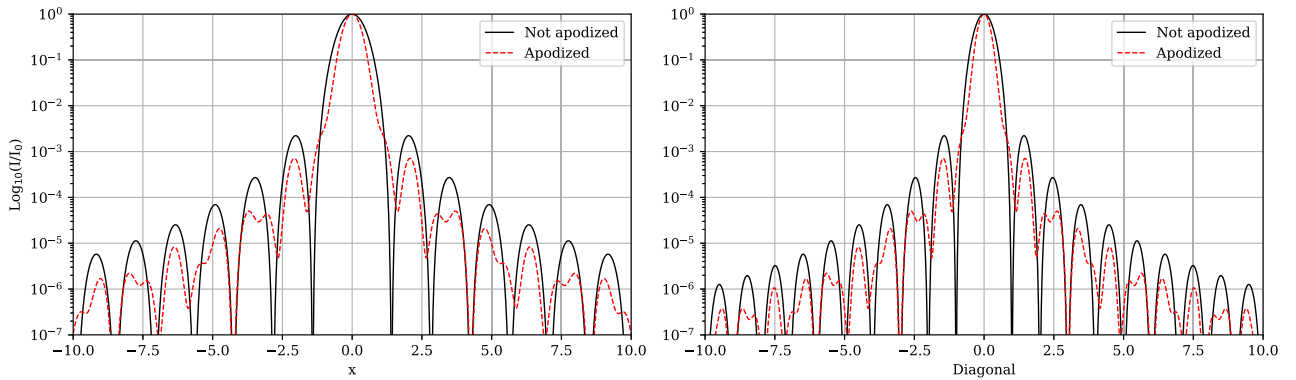

**Figure 8.** The logarithmic radial cuts of the PSF intensity for a rectangular aperture are shown along the *x*-axis (left) and the diagonal cut $x = y$ (right). The unapodized case is represented by the black line, while the red line represents the apodized PSF intensity with optimal values of $\gamma_{max} = 1.928$ and $\eta_{max} = 1.657$.

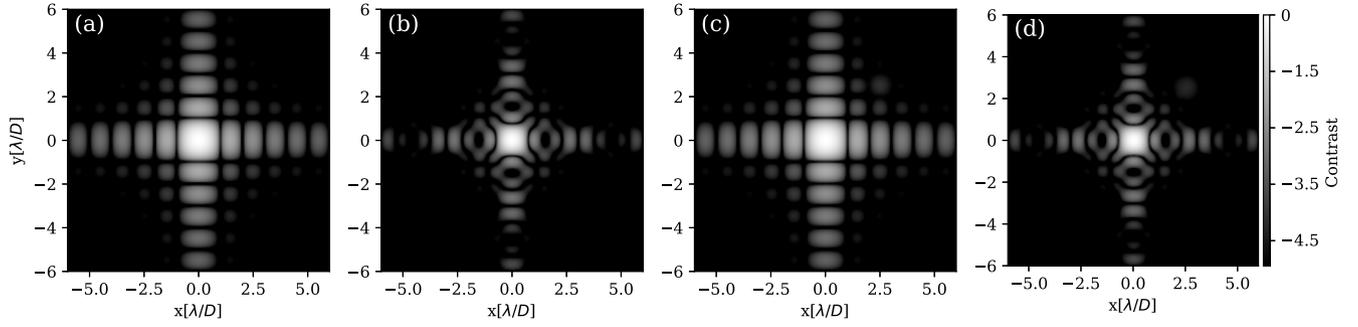

**Figure 9.** The normalized 2D intensity is shown in the figure. Panel (a) displays the intensity of the star without apodization, while panel (b) shows the intensity of the star with apodization using the optimal values obtained through optimization ($\gamma_{max} = 1.928$; $\eta_{max} = 1.657$). Panels (c) and (d) represent the intensity of the star and the planet without and with apodization, respectively, with an intensity ratio (star to planet) on the order of $10^{-4.5}$ at an angular distance of $2.5\lambda/D$. The colour bar indicates the contrast level for all images (a, b, c, and d).

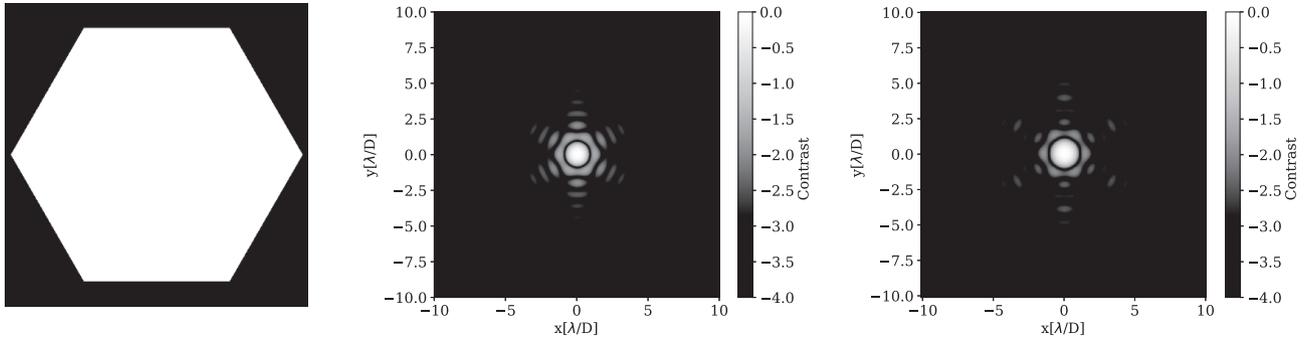

**Figure 10.** A simple hexagonal pupil was simulated, with a depiction shown on the left. The central area of the figure displays the PSF (in intensity) corresponding to the unapodized aperture. On the right side of the figure, the PSF (in intensity) is depicted for the apodized case.

### 5.1 Simple and obstructed hexagonal

The primary objective of this study was to assess the impact of IAH apodization on hexagonal apertures, which are commonly used in current and future telescopes. Regarding the simple hexagonal aperture, Fig. 10 (*left*) illustrates a simple hexagonal aperture, and the corresponding PSFs are displayed in the center of the figure for the unapodized case and on the right for the apodized case. The influence of IAH is clearly observable in the apodized case, resulting in a significant reduction of diffraction lobes from the hexagonal aperture. Moreover, significant attenuation can be noticed in the diagonal areas compared to the unapodized case. To improve the representation of the results, intensity profiles were obtained along the bright axis, as illustrated in Fig. 12 (*left*). The dashed blue curve symbolizes the case without apodization, while the red curve represents the apodized case, all on a logarithmic scale. Thanks to IAH apodization, it was easy to achieve a contrast of $10^{-6}$ at a distance of 2.1 $\lambda/D$, and a contrast of the order of $10^{-5}$ and $10^{-8}$ for angular separations ranging from 3 $\lambda/D$ to 10 $\lambda/D$.

To evaluate the impact of obstruction on the final image and the effectiveness of the IAH technique for this type of aperture, which is widely used in large observatories, we conducted simulations using an obstructed hexagonal aperture with an obstruction representing 5 per cent of the total aperture. The aperture is depicted in Fig. 11, where the left-hand panel displays the obstructed hexagonal pupil. Our simulations demonstrate the capability of IAH to significantly reduce diffraction effects, as shown in the center and right-hand panels of Fig. 11, with and without apodization, respectively. Intensity profiles obtained along the bright axis of the PSFs are presented in Fig. 12 (*right*), where the dashed blue curve corresponds to the case without apodization and the red curve corresponds to the apodized case. All curves are shown on a logarithmic scale. The results show that an average contrast of the order of $10^{-3.8}$ can be obtained at a distance between $1.7\lambda/D$ and $2.5\lambda/D$, one can also observe a contrast greater than $10^{-6}$ at a distance of $7.8\lambda/D$. Moreover, the performance achieved with our IAH approach is comparable to that of other apodization techniques aimed at mitigating the effects of stellar light diffraction. For instance, Leboulleux, Carlotti & N'Diaye (2022) utilized a segment-based optimization method to enhance the alignment robustness of a telescope with apodized segments. These findings confirm the effectiveness of IAH in improving the performance of coronagraphs, irrespective of the geometric shape of the telescope pupil.

### 5.2 TMT Aperture

We have applied the IAH to a segmented aperture similar to that used in the Thirty Meter Telescope (TMT; Nelson & Sanders 2008; Schöck et al. 2009; Cole 2017), one of the largest and most technologically advanced telescopes in the world. Scheduled for construction at Mauna Kea, Hawaii, USA, and it is planned that the telescope will have its first light in the early 2030s, this ground-based astronomical observatory is a collaborative effort among astronomical communities from India, Canada, China, Japan, and the United States, who are jointly determining the scientific goals, array of instruments, and overall design of the TMT observatory (Sivarani et al. 2022). The observatory boasts a 30-m diameter segmented mirror, comprising 492 hexagonal elements, each with a corner-to-corner measurement of 1.44 m. These segments are meticulously arranged, maintaining a mere 2.5 mm gap between them (Nelson & Sanders 2006). The primary segmented design offers several advantages, including the reduction of many difficulties associated with the construction of large telescopes, such as manufacturing, testing, risk mitigation, and transportation of large mirrors and mirror cells. Additionally, the need for significant handling equipment, large-capacity handling cranes, and large vacuum coating chambers is also reduced.

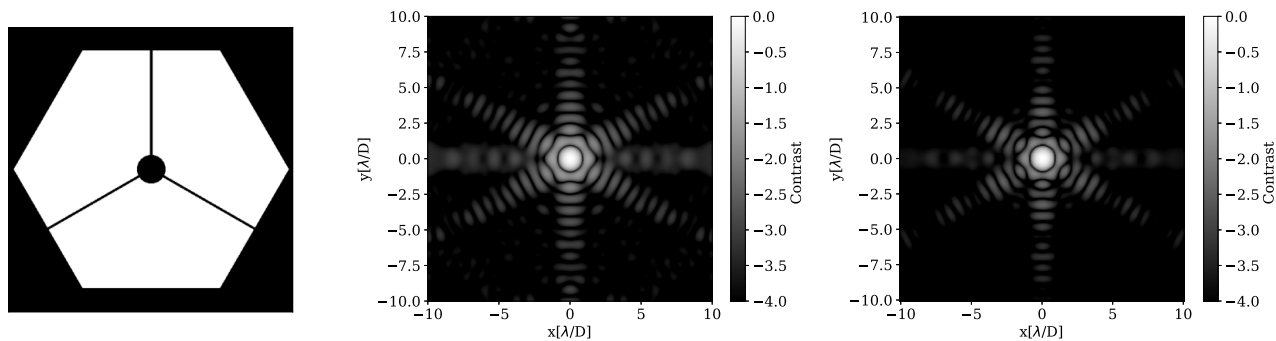

**Figure 11.** Obstructed hexagonal pupil and displayed it on the left-hand side of the figure. In the central area of the figure, PSF (in intensity) corresponding to the aperture without apodization. On the right-hand side of the figure, PSF (in intensity) for the apodized case.

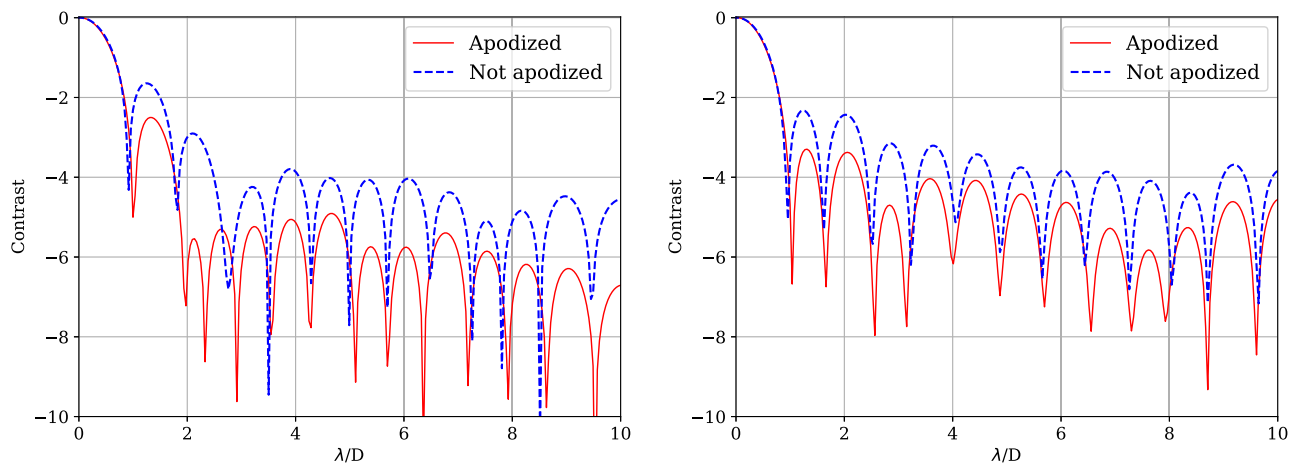

**Figure 12.** Intensity profiles along the bright axis of the PSFs for the case of a simple hexagonal aperture (left) and a hexagonal aperture with central obstruction (right) are shown. The dotted blue curves represent the case without apodization, and the red curves represent the apodized case. All curves are displayed on a logarithmic scale.

This subsection outlines the findings from simulating interferometric apodization through homothety on an aperture comparable to those employed in the TMT. The objective of this research is to assess the impact of IAH on segmented apertures frequently found in modern and forthcoming large telescopes (e.g. TMT, VLT, ELT, etc.). Fig. 13 illustrates the telescope aperture on the left, the intensity (PSF) without apodization in the center, and the intensity with apodization on the right. The impact of IAH apodization is readily apparent, especially in the diagonal regions when compared to the unapodized case. To depict the apodization effect, we computed the azimuthal average, which is displayed in Fig. 14. The curves are displayed normalized and on a logarithmic scale, with the blue curve representing the case without apodization and the red curve indicating the apodized case. Based on these simulation results, we observed that the IAH technique achieves a contrast on the order of $10^{-5}$ at a distance of $2\lambda/D$, as well as an average contrast of $5 \times 10^{-6}$ at angular separations between $4\lambda/D$ and $10\lambda/D$, we can also observe a highly significant contrast of $10^{-7}$ at an exact distance of $8.2\lambda/D$. The research conducted by Anche et al. (2023) reveals that coronagraphs for $R$ and $I$ bands achieve a contrast greater than $10^{-4}$ for ELT and TMT, but only greater than $10^{-5}$ for GMT. This implies that significant improvements must be made to coronagraphs, using apodization techniques such as IAH, to achieve the necessary contrast levels for high-contrast imaging instruments. A post-processing contrast of $10^{-8}$ at $2\lambda/D$ is expected to be achieved. To achieve this level of performance, a raw contrast of $10^{-5}$ to $1\lambda/D$ at $2\lambda/D$ is required, as noted by Fitzgerald et al. (2019). Given that IAH technology can achieve a contrast of $10^{-6}$–$10^{-7}$, its crucial role in achieving the objective of the TMT project is evident. Apodization IAH is a highly effective method for mitigating the diffraction effects of stellar light caused by the segmented aperture's shape. We chose this aperture style because its large number of segments makes it the most complex, resulting in significant diffraction effects.

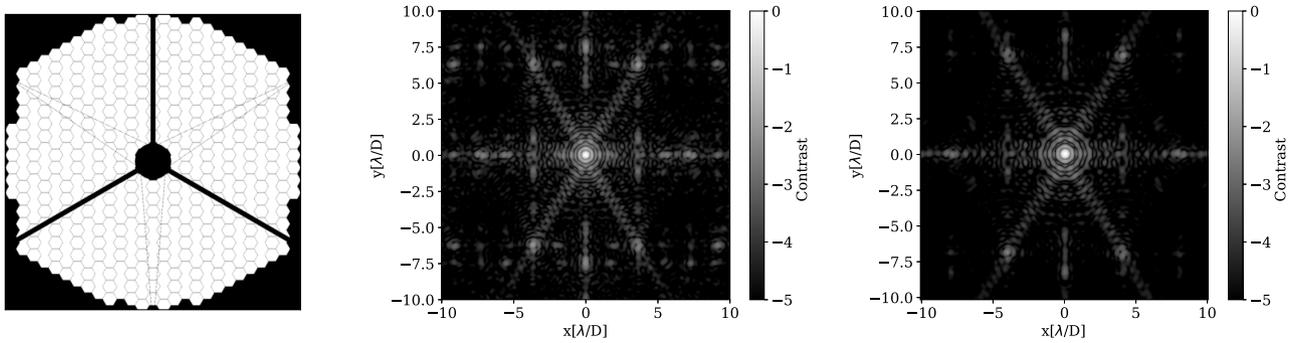

**Figure 13.** Segmented aperture, similar to that used in the TMT telescope and shown on the left side. In the central area, the PSF (in intensity) corresponds to the aperture without apodization. On the right, the PSF (in intensity) is shown for the apodized case.

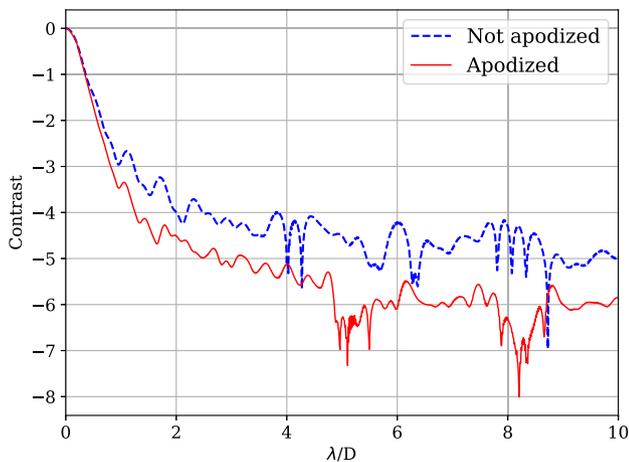

**Figure 14.** Normalized azimuthal mean of the PSF (in intensity), presented on a logarithmic scale for a segmented aperture similar to that planned for the TMT telescope. The blue curves represent the non-apodized case, while the red curves correspond to the apodized case.

IAH successfully reduces these effects considerably, leading us to conclude that IAH can be applied to any geometric shape of the telescope's pupil.

## 6 CONCLUSION

In this work, we presented a method for interferometric apodization of circular, rectangular or hexagonal apertures using homothety. We optimized the free parameters of the system to concentrate the maximum luminous energy from the star in the central lobe of diffraction. Our results showed that the total luminous energy is concentrated in the central disc, with a percentage of 93.6 per cent for the circular geometry, and 91.5 per cent for the rectangular aperture. This is a significant improvement compared to the non-apodized case where the percentage of light intensity contained in the main peak is only 81.5 per cent. We also presented standard values ($\gamma = 0.498$; $\eta = 0.6$) for the practical set-up, which were found to be slightly less optimal than the optimal values ($\gamma = 0.525$; $\eta = 0.625$). We demonstrated through simulations for both circular and rectangular apertures that the optimized IAH method can improve the detection of exoplanets with an intensity ratio (star/exoplanet) of the order of $10^{-4}$ from $1.8\lambda/D$. Our simulations confirm the efficacy of IAH in improving coronagraph performance, irrespective of the telescope pupil's geometric shape. Hexagonal apertures are increasingly being adopted in large aperture telescopes due to their technologically viable solution to the challenging problem of constructing telescopes with ever larger primary mirrors (Sabatke et al. 2005; Soummer et al. 2009). The azimuthal averages of the PSFs provide a more accurate representation of the results. These findings hold significance in guiding the development of advanced coronagraphs. We conclude that IAH is a promising technique for improving coronagraph performance, irrespective of the telescope pupil's geometric shape. For our forthcoming work, which will be detailed in the upcoming article, we plan to present a comparison of simulation predictions with experimental results, including a systematic examination of the effect of various parameters.

## ACKNOWLEDGEMENTS

We would like to extend our heartfelt appreciation to Y. Attaourti and Y. Moulane for graciously devoting their time to review this work. Their unwavering support and invaluable guidance throughout this process are deeply appreciated. We would like to express our sincere gratitude to M. N'Diaye for his valuable contributions to this research project. His insightful questions, comments, and expertise have greatly enhanced the quality of our work.

## DATA AVAILABILITY

No data is available for sharing. It is solely based on simulation.

*Interferometric apodization by homothety* 5451Belikov R. et al., 2022, in Coyle L. E., Matsuura S., Perrin M. D.eds, Proc. SPIE Conf. Ser. Vol. 12180, Space Telescopes and Instrumentation 2022: Optical, Infrared, and Millimeter Wave. SPIE, Bellingham, p. 1218025

Born M., Wolf E., 1980, Principles of Optics Electromagnetic Theory of Propagation, Interference and Diffraction of Light. GB Pergamon Press, Oxford, Great Britain

Bos S. P. et al., 2021, A&A, 653, A42

Carlotti A., Ricort G., Aime C., El Azhari Y., Soummer R., 2008, A&A, 477, 329

Cole G., 2017, in Optical Design and Fabrication 2017 (Freeform, IODC, OFT). Optica Publishing Group, Denver, Colorado, United States, p. OW1B.4

El Azhari Y., Azagrouze O., Martin F., Soummer R., Aime C., 2006, in Aime C., Vakili F.eds, Proc. IAU Colloq. 200. Direct Imaging of Exoplanets: Science and Techniques. Cambridge Univ. Press, Cambridge, p. 445

Fitzgerald M. et al., 2019, in Bulletin of the American Astronomical Society. AAS, United States, p. 251

Fogarty K., Pueyo L., Mazoyer J., N'Diaye M., 2017, AJ, 154, 240

Foo G., Palacios D. M., Swartzlander Grover A. J., 2005, Opt. Lett., 30, 3308

Guyon O., Martinache F., Belikov R., Soummer R., 2010, ApJS, 190, 220

Habib A., Azagrouze O., El Azhari Y., Benkhaldoun Z., Lazrek M., 2010, MNRAS, 406, 2743

Itoh S., Matsuo T., 2022, AJ, 163, 279

Juanola-Parramon R. et al., 2022, J. Astron. Telesc. Instrum. Sys., 8, 034001

Laginja I. et al., 2019, in Proc. SPIE Conf. Ser. Vol. 11117, Techniques and Instrumentation for Detection of Exoplanets IX. SPIE, Bellingham, p. 1111717

Laginja I. et al., 2020, in Proc. SPIE Conf. Ser. Vol. 11443, Space Telescopes and Instrumentation 2020: Optical, Infrared, and Millimeter Wave. SPIE, Bellingham, p. 114433J

Laginja I., Soummer R., Mugnier L. M., Pueyo L., Sauvage J.-F., Leboulleux L., Coyle L., Knight J. S., 2021, J. Astron. Telesc. Instrum. Syst., 7, 015004

Leboulleux L., Sauvage J.-F., Pueyo L., Fusco T., Soummer R., N'Diaye M., St. Laurent K., 2017, in Proc. SPIE Conf. Ser. Vol. 10400, Techniques and Instrumentation for Detection of Exoplanets VIII. SPIE, Bellingham, p. 104000M

Leboulleux L. et al., 2018a, J. Astron. Telesc. Instrum. Syst., 4, 035002

Leboulleux L., Pueyo L., Sauvage J.-F., Fusco T., Mazoyer J., Sivaramakrishnan A., N'Diaye M., Soummer R., 2018b, in Lystrup M., MacEwen H. A., Fazio G. G., Batalha N., Siegler N., Tong E. C.eds, Proc. SPIE Conf. Ser. Vol. 10698, Space Telescopes and Instrumentation 2018: Optical, Infrared, and Millimeter Wave. SPIE, Bellingham, p. 106986H

Leboulleux L., Carlotti A., N'Diaye M., 2022, A&A, 659, A143

Marx D., Belikov R., Sirbu D., Kern B., Prada C., Bendek E., Wilson D., 2021, in Shaklan S. B., Ruane G. J.eds, Proc. SPIE Conf. Ser. Vol. 11823, Techniques and Instrumentation for Detection of Exoplanets X. SPIE, Bellingham, p. 118230O

Mawet D., Riaud P., Absil O., Surdej J., 2005, ApJ, 633, 1191

N'Diaye M., Pueyo L., Soummer R., 2015, ApJ, 799, 225

N'Diaye M., Soummer R., Pueyo L., Carlotti A., Stark C. C., Perrin M. D., 2016, ApJ, 818, 163

N'Diaye M. et al., 2018, in Lystrup M., MacEwen H. A., Fazio G. G., Batalha N., Siegler N., Tong E. C.eds, Proc. SPIE Conf. Ser. Vol. 10698, Space Telescopes and Instrumentation 2018: Optical, Infrared, and Millimeter Wave. SPIE, Bellingham, p. 106986A

Nelson J., Sanders G. H., 2006, in Stepp L. M.ed, Proc. SPIE Conf. Ser. Vol. 6267, Ground-based and Airborne Telescopes. SPIE, Bellingham, p. 626728

Nelson J., Sanders G. H., 2008, in Stepp L. M., Gilmozzi R.eds, Proc. SPIE Conf. Ser. Vol. 7012, Ground-based and Airborne Telescopes II. SPIE, Bellingham, p. 70121A

Nickson B. F. et al., 2022, in Coyle L. E., Matsuura S., Perrin M. D.eds, Proc. SPIE Conf. Ser. Vol. 12180, Space Telescopes and Instrumentation 2022: Optical, Infrared, and Millimeter Wave. SPIE, Bellingham, p. 121805K

Nisenson P., Papaliolios C., 2001, ApJ, 548, L201

Otten G. P. P. L., Snik F., Kenworthy M. A., Miskiewicz M. N., Escuti M. J., 2014, Opt. Express, 22, 30287

Por E. H., 2017, in Shaklan S.ed, Proc. SPIE Conf. Ser. Vol. 10400, Techniques and Instrumentation for Detection of Exoplanets VIII. SPIE, Bellingham, p. 104000V

Por E. H., 2020, ApJ, 888, 127

Por E. H., Soummer R., Noss J., St. Laurent K., 2020, in Proc. SPIE Conf. Ser. Vol. 11443, Space Telescopes and Instrumentation 2020: Optical, Infrared, and Millimeter Wave. SPIE, Bellingham, p. 114433P

Por E. H., Pourcelot R., Klein A., Soummer R., Pueyo L., Sahoo A., Nguyen M., Fox R., 2023, in American Astronomical Society Meeting Abstracts. p. 461.08

Ridgway S. T., 2004, in Aime C., Soummer R.eds, EAS Publ. Ser. Vol. 12, Astronomy with High Contrast Imaging II. Cambridge University Press, Cambridge, p. 49

Ruane G., Mawet D., Mennesson B., Jewell J., Shaklan S., 2018, J. Astron. Telesc. Instrum. Syst., 4, 015004

Sabatke E., Burge J., Sabatke D., 2005, Appl. Opt., 44, 1360

Schöck M. et al., 2009, PASP, 121, 384

Sirbu D. et al., 2021, in Shaklan S. B., Ruane G. J.eds, Proc. SPIE Conf. Ser. Vol. 11823, Techniques and Instrumentation for Detection of Exoplanets X. SPIE, Bellingham. p. 118230R

Sivarani T. et al., 2022, J. Astrophys. Astron., 43, 86

Snik F., Otten G., Kenworthy M., Miskiewicz M., Escuti M., Packham C., Codona J., 2012, in Navarro R., Cunningham C. R., Prieto E.eds, Proc. SPIE Conf. Ser. Vol. 8450, Modern Technologies in Space- and Ground-based Telescopes and Instrumentation II. SPIE, Bellingham, p. 84500M

Soummer R., 2005, ApJ, 618, L161

Soummer R., Aime C., Falloon P. E., 2003, A&A, 397, 1161

Soummer R., Pueyo L., Ferrari A., Aime C., Sivaramakrishnan A., Yaitskova N., 2009, ApJ, 695, 695

Soummer R., Sivaramakrishnan A., Pueyo L., Macintosh B., Oppenheimer B. R., 2011, ApJ, 729, 144

Stahl H. P., 2017, in Proc. SPIE Conf. Ser. Vol. 10398, UV/Optical/IR Space Telescopes and Instruments: Innovative Technologies and Concepts VIII. SPIE, Bellingham, p. 1039806

St. Laurent K. et al., 2018, in Lystrup M., MacEwen H. A., Fazio G. G., Batalha N., Siegler N., Tong E. C.eds, Proc. SPIE Conf. Ser. Vol. 10698, Space Telescopes and Instrumentation 2018: Optical, Infrared, and Millimeter Wave. SPIE, Bellingham, p. 106982W

St. Laurent K. et al., 2019, in American Astronomical Society Meeting Abstracts #233. AAS, Washington, United States, p. 157.40

Stahl M., Shaklan S., Stahl H. P., 2015, in Shaklan S.ed, Proc. SPIE Conf. Ser. Vol. 9605, Techniques and Instrumentation for Detection of Exoplanets VII. SPIE, Bellingham, p. 96050P

Stahl H. P., Nemati B., Stahl M. T., 2020, in Modeling, Systems Engineering, and Project Management for Astronomy IX. SPIE, p. 114502Z

The LUVOIR Team, 2019, preprint (arXiv:1912.06219)

Will S. D., Fienup J. R., 2020, J. Opt. Soc. Am. A, 37, 629

Zanoni C. A., Hill H. A., 1965, J. Opt. Soc. Am., 55, 1608

Zhang M., Ruane G., Delorme J.-R., Mawet D., Jovanovic N., Jewell J., Shaklan S., Wallace J. K., 2018, in Lystrup M., MacEwen H. A., Fazio G. G., Batalha N., Siegler N., Tong E. C.eds, Proc. SPIE Conf. Ser. Vol. 10698, Space Telescopes and Instrumentation 2018: Optical, Infrared, and Millimeter Wave, SPIE, Bellingham, p. 106985X
This paper has been typeset from a TeX/LaTeX file prepared by the author.

© 2023 The Author(s)
Published by Oxford University Press on behalf of Royal Astronomical SocietyMNRAS **523**, 5442–5451 (2023)